\documentclass[journal]{IEEEtran}
\usepackage{hyperref}
\usepackage{lettrine}
\usepackage{epsfig}
\usepackage[linesnumbered,boxed,ruled,commentsnumbered]{algorithm2e}
\usepackage{footnote}
\usepackage{cite}
\usepackage{float}
\usepackage{booktabs}
\usepackage{multicol}
\usepackage{dblfloatfix}
\usepackage{amssymb}
\usepackage{amsmath}
\usepackage{amsthm}
\usepackage{comment}
\usepackage{lscape}
\usepackage{multirow}
\usepackage{amsmath,mathabx}
\usepackage{color}
\usepackage{amsfonts}
\usepackage{tabularx,tabulary}
\usepackage[font=footnotesize]{caption}
\usepackage{graphicx}
\usepackage{bbm}
\usepackage{bigints}
\usepackage[usestackEOL]{stackengine}
\usepackage{color}
\usepackage{multicol}
\usepackage{multirow}
\usepackage{caption}
\usepackage{subcaption}
\usepackage{tabularx}
\usepackage[T1]{fontenc}
\usepackage[shortlabels]{enumitem}

\usepackage{verbatim} 

\usepackage[pagewise]{lineno}

\usepackage{scalerel}
\usepackage{tikz}
\usetikzlibrary{svg.path}
\definecolor{orcidlogocol}{HTML}{A6CE39}
\tikzset{
  orcidlogo/.pic={
    \fill[orcidlogocol] svg{M256,128c0,70.7-57.3,128-128,128C57.3,256,0,198.7,0,128C0,57.3,57.3,0,128,0C198.7,0,256,57.3,256,128z};
    \fill[white] svg{M86.3,186.2H70.9V79.1h15.4v48.4V186.2z}
                 svg{M108.9,79.1h41.6c39.6,0,57,28.3,57,53.6c0,27.5-21.5,53.6-56.8,53.6h-41.8V79.1z M124.3,172.4h24.5c34.9,0,42.9-26.5,42.9-39.7c0-21.5-13.7-39.7-43.7-39.7h-23.7V172.4z}
                 svg{M88.7,56.8c0,5.5-4.5,10.1-10.1,10.1c-5.6,0-10.1-4.6-10.1-10.1c0-5.6,4.5-10.1,10.1-10.1C84.2,46.7,88.7,51.3,88.7,56.8z};
  }
}

\newcommand\orcidicon[1]{\href{https://orcid.org/#1}{\mbox{\scalerel*{
\begin{tikzpicture}[yscale=-1,transform shape]
\pic{orcidlogo};
\end{tikzpicture}
}{|}}}}

\begin{document}

\title{ How to Leverage High Altitude Platforms in Green Computing?  }

\author{Wiem Abderrahim \orcidicon{0000-0001-5896-2307}, Osama Amin \orcidicon{0000-0002-0026-5960} and Basem Shihada \orcidicon{0000-0003-4434-4334}\\
\thanks{The authors are with the  Computer, Electrical and Mathematical Sciences and Engineering (CEMSE) Divison, King Abdullah University of Science and Technology (KAUST), Thuwal 23955, Makkah Prov., Saudi Arabia (e-mail: { wiem.abderrahim, osama.amin, basem.shihada }@kaust.edu.sa)}}

\maketitle

\begin{abstract}
Terrestrial data centers suffer from a growing carbon footprint that could contribute with $14\%$ to global CO2 emissions by 2040.
High Altitude Platform (HAP) is a promising airborne technology that can unleash the computing frontier in the stratospheric range by hosting a flying data center. HAP systems can endorse the sustainable green operation of data centers thanks to the naturally low atmospheric temperature that saves cooling energy and its large surface that can host solar panels covering energy requirements. Throughout this article, we define the operation limitations of this innovative solution and study the energy-efficiency-related trade-offs. Then, we shed light on the significance of the scalability of the data center-enabled HAP architecture by investigating potential bottlenecks and proposing different deployment scenarios to avoid network congestion. We also highlight the importance of the management agility of the data center-enabled HAP system by defining effective management techniques that yield high-performing data centers. Our results demonstrate that deploying a single data center-enabled HAP can save $12\%$ of the electricity costs.

\end{abstract}

\section{Introduction}

Climate change is a momentous threat that humankind faces because it  severely affects the health of the planet and jeopardizes all forms of  life. Therefore, prompt and collective actions should be taken to decelerate its effects by reducing excessive carbon emissions worldwide. One of the major industries that is raising serious concerns in this regard is the data center industry; which is anticipated to  contribute with $14\%$ to global CO2 emissions  by 2040 \cite{yang2021carbon}. Indeed, data centers will consume  around $13\%$ of the worldwide electricity by 2030; which is  predominately generated by carbon-intensive fossil fuels  \cite{koronen2020data}.  Moreover, these large-scale computing infrastructures; which are already behind major energy-efficiency issues, are expected to expand exponentially to process and store our continuously-growing data generated by data-intensive applications emerging in different fields such as  artificial intelligence, smart cities and telecommuting \cite{dukic2020beyond}. For example, the total number of installed instances such as virtual machines and containers in global data centers increased from about 150 million in 2016 to 500 million in 2021 \cite{lin2020disaggregated}. Moreover, Amazon has deployed large-scale data centers in 25 geographic regions to support 80 availability zones   and Google has deployed more than 18 data centers in Americas, Asia, and Europe recently \cite{yang2021carbon}. Given this alarming situation, joint efforts from academia and industry  are required  to improve the  energy-efficiency of data centers  beyond recommending the commonly adopted approaches that are based on renewable energy usage  or  carbon offset mechanisms \cite{cao2022towards}. 

Pragmatically, non-traditional computation paradigms are needed to provide innovative and tangible solutions that vigorously tackle the energy-efficiency issues of current data centers. Different technologies can be incorporated with the terrestrial data center to contribute to environmental sustainability by providing carbon-neutral integrated data centers. {Underwater data centers are currently publicized as the future of data storage thanks to their high reliability}. Moreover,  satellite data centers have attracted growing interest recently as the newly defined computing frontier. However, most of these data center's architectures present several drawbacks ranging from limited coverage for underwater data centers to high costs, additional delays, and limited payload for satellite data centers. Interestingly, high-altitude platforms (HAP) seem a good trade-off between these incorporated technologies because they offer larger coverage areas than underwater systems, support more important payloads, and guarantee easier maintenance and lower delays than satellites.     

 \begin{figure*}[h]
      \centering
      \includegraphics[scale=0.3]{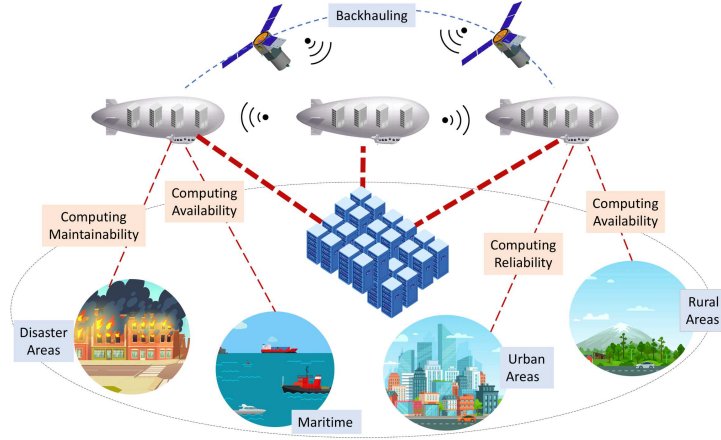}
      \caption{Data Center Enabled HAP Use Cases}
      \label{fig:fig1}
  \end{figure*}

Therefore, we believe that the data center-enabled High Altitude Platform (HAP)  is a distinguishable green alternative to traditional terrestrial data centers, given the encouraging merits of the HAPs. Indeed, HAP can be a core futuristic airborne network component that will unleash the networking frontier in the stratospheric range at an altitude between 17 km and 20 km \cite{HAPEnabl,kurt2021vision}. HAP offers several unique advantages, especially from energy and communication perspectives. Firstly, the HAP location at the stratosphere saves the cooling energy thanks to the naturally low atmospheric temperature (between $-50^{\circ}\text{C}$ and $-15^{\circ}\text{C}$). Hence, a data center-enabled HAP  can offload some workload from terrestrial data centers, saving the associated cooling energy. Moreover, HAP can host large solar panels that harvest substantial amounts of solar energy thanks to HAP's large surface and its location above the clouds. {The HAP feeds the servers with the solar energy harvested during the daytime and stored in the Lithium-Sulphur batteries during the nighttime. Hence, the harvested solar energy can cover amply the computational power required by the data center' servers; while the necessary energy conversion and management strategies are efficiently applied \cite{kurt2021vision, HAPEnabl}}. {Secondly, the HAP's location at high altitudes offers LoS communication links with several receivers thanks to the the availability of large terrestrial footprint and the absence of obstructions in the horizon. Hence, HAPs can establish reliable direct links with a large number of terrestrial base stations \cite{kurt2021vision, HAPEnabl}}. These advantages enable the data center-enabled HAP to offer a rich panoply of computing services that range from supporting internet of things  applications to intelligent transportation systems  as depicted in Fig~.\ref{fig:fig1}. Moreover, the  data center-enabled HAP improves the  dependability properties of these computing services by boosting the reliability of the terrestrial data center in urban areas and its maintainability in disaster areas. Besides, the flying data center hosted in the HAP guarantees the availability of the computing services  in rural and remote areas in addition to under-connected areas such as  maritime area.

In this work, we study the data center-enabled HAP architecture from different aspects related to energy-awareness, scalability and management agility while defining the limits and conditions under which this solution can be leveraged for green computing. First, we investigate the major settings and trade-offs that impact the performance of the data center-enabled HAP  in terms of  energy saving.  Then, we analyse the potential bottlenecks of this architecture and we provide various deployment scenarios to circumvent the anticipated scalability issues. Finally, we advocate different management techniques to preserve the desired operation of the data center-enabled HAP  over long-duration missions.

\section{Energy-awareness of Data center-enabled HAP}
In this section, we study the data center-enabled HAP from an energy-awareness perspective. Specifically, we analyze the conditions and trade-offs that impact the energy-saving benefits of the data center-enabled HAP.  {We conduct all the simulations based on realistic parameters detailed in Table I.}

\begin{table}[h]
\caption{ Simulation Settings}
\centering {
\begin{tabular}{|c|c |c|} 
 \hline
Type &Parameter & Numerical Value  \\ 
 \hline\hline
& Supply Temperature  & $299.15$ K\\
Cooling/& Server Initial Temperature  & $310$ K\\
Thermal &CPU Initial Temperature  & $318$ K\\
Inputs & Thermal Resistance & $0.34$ K/W\\
 & Server Heat Capacity  & $340$  J/K\\
 \hline
  & Maximum Payload  & $450\:\mathrm{kg}$ \\
  & Flying Server Payload   & $11\:\mathrm{kg}$ \\
  HAP & Area of the Photovoltaic (PV)  Surface   & $8000\:\mathrm{m}^2$ \\
Inputs &Efficiency of the PV   & $0.4$ \\
 &Propeller efficiency  & $0.8$ \\
& Battery capacity  & $2\:\mathrm{kWh/kg}$ \\
 \hline
 Trans-& Antennas in TDC  & 2  \\
mission &Antennas  in HAP  & 16  \\
Inputs &Carrier Frequency  & $31$ GHz \\
&Channel Bandwidth   & $100$ MHz \\
 \hline 
\end{tabular}}
\label{table:1}
\end{table}

\subsection{Energy Saving} 
A breakdown of the energy consumed by a data center shows that the cooling infrastructure  and the computational infrastructure   are the main components that absorb the data center energy with up to $40\%$ and  $56\%$ respectively \cite{khalid2022dual}. Therefore, the data center-enabled HAP can significantly reduce the energy consumed by a terrestrial data center  thanks to two main reasons. First, the data center-enabled HAP saves the cooling energy thanks to the naturally low atmospheric temperature of the stratosphere (between $-50^{\circ}\text{C}$ and $-15^{\circ}\text{C}$); where the HAP is located. This advantage dismisses the need for the presence of cooling units in the HAP because the average temperature in the stratosphere is substantially lower than the recommended temperature for the data center (between $18^{\circ}{C}$ and $26^{\circ}\text{C}$) by the American Society of Heating, Refrigerating and Air-Conditioning Engineers. Second, the data center-enabled HAP saves the computational energy thanks to the harvested solar energy. Indeed, the HAP feeds the servers with the solar energy harvested during the daytime and stored in the Lithium-Sulphur batteries during the nighttime. However, the terrestrial data center feeds its servers with the electric energy supplied through the electrical grid constantly.

We note that the energy saved by the data center-enabled HAP is impacted by the HAP's location and the considered period of the year. For instance, the HAP can harvest more solar energy during June if it is located in the Northern hemisphere. Indeed, the solar radiation and the daylight duration are more critical since the Northern hemisphere is closer to the Sun in June.  

We also highlight the vital role of distributed learning in improving the energy efficiency of the data center-enabled HAP. Indeed, the optimization of such complex non-linear systems geographically distributed between the sky and the ground by conventional optimization frameworks requires complicated heuristics and tedious calculations. However, implementing a distributed learning algorithm that trains the data of the airborne servers and the terrestrial servers yields a global energy-efficiency model that predicts the necessary policies with less required complexity.  {Since learning approaches are generally energy-hungry, the adopted algorithms should be carefully selected to conserve the energy-efficiency of the data center-enabled HAP \cite{barua2022green}}.




\subsection{HAP Flying Condition} 
 Energy management of HAPs is crucial because these platforms are typically designed for long-duration missions. {Therefore, we establish the HAP flying condition to examine when the HAP can stay stable and keep flying. Hence, we study the energy budget of the HAP such that the daily harvested energy is  equal to the consumed energy. The main energy source for the daytime operation of HAPs is the solar energy harvested  {through the photovoltaic system implemented on the surface of the inflatable part of the HAP. }The harvested energy depends mainly on the latitude of the HAP and the considered day of the year \cite{kurt2021vision,HAPEnabl,arum2020energy}.} HAPs also incorporate energy storage components; which are typically Lithium-Sulphur batteries or hydrogen fuel cells.  These batteries support the nighttime operation of the HAP and are fed by the solar energy harvested during the daytime. {Under the flying condition, the harvested energy should cover the consumed energy by the HAP; which captures the energy consumed by its propulsion subsystem, the energy consumed by its payload subsystem and wireless communication energy. The propulsion energy includes the energy required to localize and stabilize the HAP. The payload energy is the computational energy of the servers hosted in the HAP and it depends on the servers characteristics (e.g. the desirable utilization ratio and the service rate in million instructions per second)  besides the workload characteristics (e.g. the arrival rate and the mean task length).}
 
\begin{figure}
    \centering
   \includegraphics[width=3.4in]{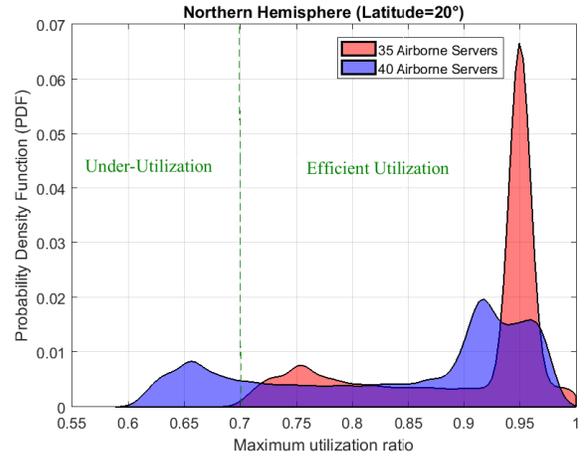}
    \label{fig:2a}
      \caption{HAP Flying Condition over the Year's Days}
      \label{fig:2}
     \end{figure}

  {To offer qualitative insight into the HAP flying condition, we study the maximum utilization ratio per server under this  condition based on realistic parameters of HAP and data centers\cite{ding2021joint,sun2020simulation}. We plot the probability distribution function  of the utilization ratio in Fig.~\ref{fig:2} over the different days of the year. Indeed, the HAP flying condition determines the maximum  computational energy consumed by the HAP-hosted servers and depends mainly on the servers' utilization ratio. We can assess the utilization ratio of a server by dividing the workload arrival rate by the server's service rate. We consider the server to be under-utilized if its utilization ratio is below $70\%$, and it is over-utilized if its utilization ratio is close to $100\%$. 
  }
   
 {The HAP flying condition is mainly impacted by the number of airborne servers hosted in the HAP.  For instance, 
  when  40 servers are deployed in the HAP, the effective usage of the airborne servers (i.e., utilization ratio between $70\%$ and $100\%$) is possible during most of the year's days as depicted in Fig.~\ref{fig:2}.  In this case, the collected solar energy covers exactly the propulsion energy and the payload energy and the servers are mostly used efficiently because the HAP is fully loaded with 40 servers.  However, when only 35 servers are deployed in the HAP,  the airborne servers need to be over-utilized (i.e., utilization ratio close to $100\%$) to fully use the harvested energy.   Therefore, the harvested energy and the servers' capacities are effectively used when the number of airborne servers increases. } 

\begin{figure*}
    \centering
  \frame{ \includegraphics[scale=0.3]{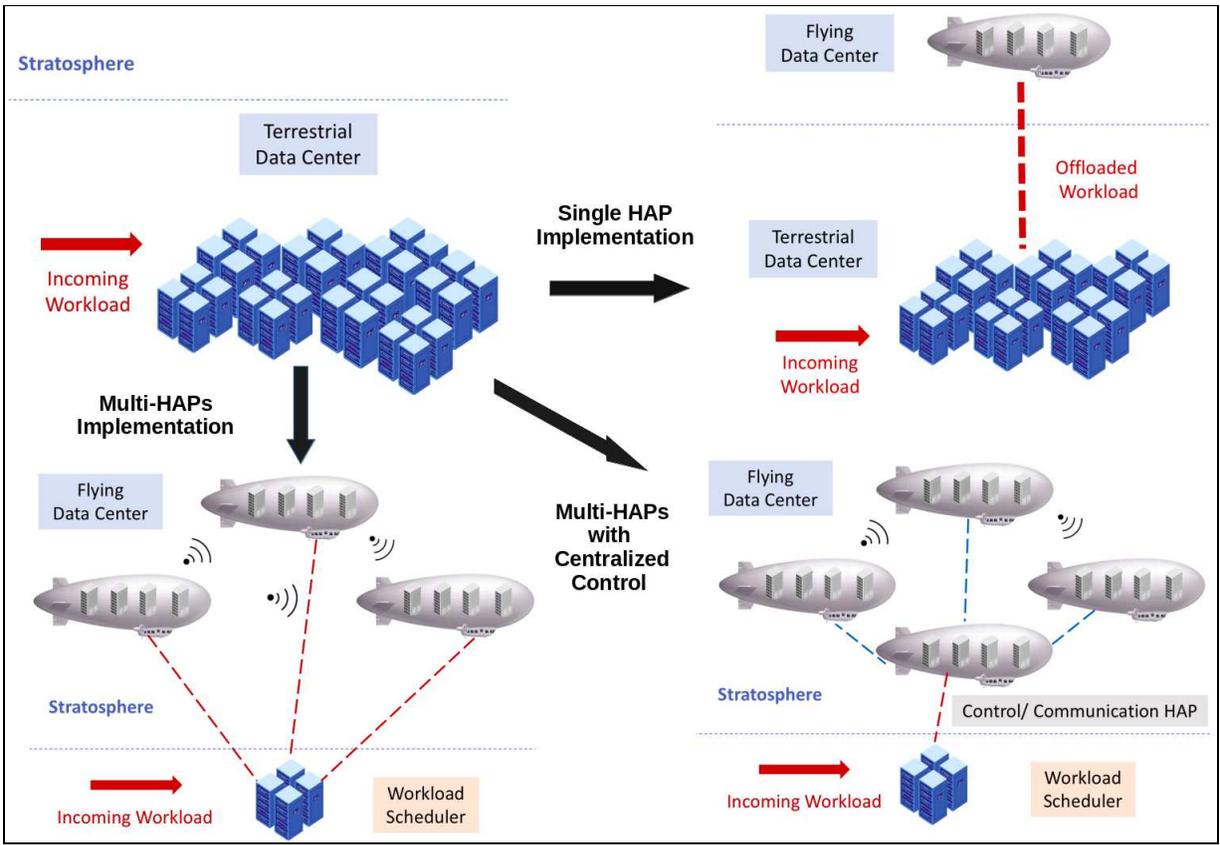}}
     \caption{Data center-enabled HAP Deployment Scenarios}
     \label{fig:3}
     \end{figure*}

\subsection{Trade-offs}
It is important to reduce the energy consumed by the terrestrial data center by maximizing the number of airborne servers and the computing workload offloaded to the HAP. However, the adoption of such techniques leads to a resource utilization dilemma since energy consumption and resource utilization are strongly coupled. On the one hand, the over-utilization of the available resources threatens the physical capabilities of the system and might yield dysfunctional servers or unbalanced HAP.
For instance, high central processing unit (CPU) utilization and/or memory utilization overload(s) the server and lead(s) to unresponsive or frozen systems. Also, high resource utilization requires high harvested energy, which may preclude the flying condition. 
On the other hand, the under-utilization of the available resources might yield aging servers and substantial wasted energy since the idle servers can consume as high as $60\%$ of the peak power \cite{khalid2022dual}. Therefore, it is valuable to adopt the adequate resource management (e.g. consolidation, containerization) techniques in data center-enabled HAP  to reduce the consumed energy without overlooking the physical capacities of the available resources.  

Another relevant trade-off to energy-efficiency is related to service availability. Indeed, data centers have strict service level agreements (SLA) that include for instance providing high availability (24/7), short delays and maximum security to the hosted services. To meet these requirements and avoid the SLA's violation penalties, resources are usually over-provisioned.  Therefore, it is valuable to find the appropriate resource provisioning and workload scheduling in the airborne servers present in the HAP that not only reduces energy consumption but also avoids performance degradation especially in terms of service continuity, reliability and security.

\section{Scalability of Data Center-Enabled HAP}
In this section, we study different deployment scenarios to overcome the scalability issues of the data center-enabled HAP architecture. 
 
   \subsection{Bottlenecks in Data Center-Enabled HAP}
According to Cisco, global cloud data center traffic was estimated to reach 20.6 zettabytes  per year in 2021 given the data-intensive applications  such as artificial intelligence and internet-of-everything  applications  \cite{kaur2019big}. This skyrocketing data consumes huge computing and communication   resources and might create  potential bottlenecks in the data center-enabled HAP. For instance, some airborne servers may become hotspots if the offloaded traffic to the HAP is not properly load-balanced. This issue leads to overloaded servers' buffers and impacts substantially the experienced queuing delay. Moreover, the wireless link between the HAP and the terrestrial data center is a typical bottleneck in the data center-enabled HAP. Indeed, link outage happens if the offloaded workload from the terrestrial data center cannot be supported by the established transmission link to the HAP. This issue leads to a supplementary outage delay besides the imposed transmission delay engendered by the servers distribution between the HAP and the terrestrial data center. It also deteriorates the system reliability and requires supplementary re-transmission to recover the lost workload. Therefore,  the servers capabilities and the link characteristics should be carefully considered when scheduling the workload to the HAP to achieve higher performance in terms of delay and reliability. Beyond conventional workload scheduling approaches, we need to develop ambitious solutions that improve the scalability of this integrated terrestrial-aerial computing system. {In this regard, we propose three  deployment scenarios for data center-enabled HAP in the next section.}

\begin{figure}
    \centering
    \includegraphics[width=3.4in]{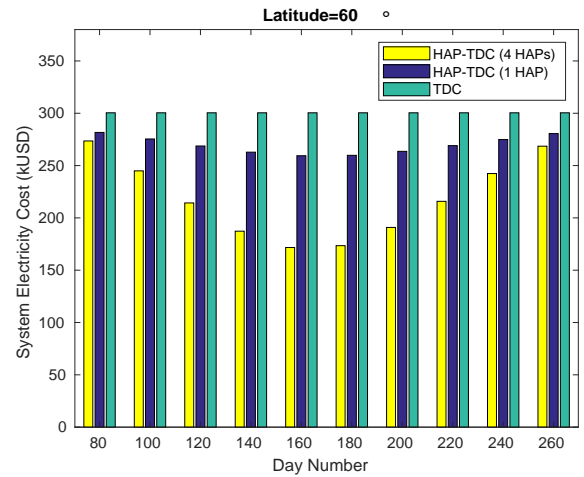}
    \caption{Year's Period Impact }
    \label{fig:4a}
\end{figure}

\subsection{Multi-HAPs Constellations}
Multi-HAPs constellations can coordinate and cooperate to support larger payloads by offering reliable services  \cite{kurt2021vision}. In our airborne computing scenario, a constellation of multi-HAPs can circumvent the  potential bottlenecks by supporting larger payloads and offering a more prominent capacity  \cite{kurt2021vision}.  {However, the server payload should be carefully considered in this configuration given its impact on the flying condition.}  Specifically, a distributed air computing platform supported by different HAPs can overcome the computational overload of one HAP. Moreover, the multi-links established between different HAPs and the terrestrial data center increase the available capacity and mitigate the link outage substantially. However, a proper constellation design is imperative to meet the growing data traffic requirements. Firstly, it is essential to determine the number of necessary airships and to optimize their locations while reducing the overlapped footprint to maximize the resources' usage.  
For instance, we should investigate when is it more effective from an energy perspective and from a performance perspective to deploy one HAP loaded with some airborne servers along with a terrestrial data center and when to deploy all the servers of the terrestrial data center in different HAPs (Fig.~\ref{fig:3}).  {Secondly, it is important to determine the interconnection type between the different HAPs (e.g. optical link, radio link) while implementing the necessary spectrum sharing and interference management techniques not only within HAPs but also with adjacent networks such as satellites' constellations and low altitude platforms  constellations. Moreover, the LOS conditions with the terrestrial data centers and between the HAPs leads to crucial interference issues  that must be mitigated through proper multiplexing techniques that meet the energy-efficiency objectives of the data center-enabled HAP.} It is  indispensable to determine if the control strategy of the multi-HAPs-enabled data center should be centralized or distributed within the HAPs \cite{alam2021high}. Specifically, in a centralized strategy (Fig.~\ref{fig:3}), the control is assigned to one of the HAPs that has the necessary information about the remaining HAPs capabilities and schedules the offloaded workload from the terrestrial data center accordingly.

In a distributed strategy, all the HAPs coordinate their communication and cooperate to manage the  offloaded workload from the terrestrial data. The centralized strategy is advantageous because the controller-HAP has a global view about the constellation and can achieve a near optimal control solution at the expense of a relatively long response time. The distributed strategy is advantageous because it circumvents the single point of failure problem at the expense of higher complexity and supplementary communication overheads. The centralized-control strategy can be improved through distributed learning approaches to optimize the control rules regarding response time and energy efficiency by offloading the appropriate workload amount to the appropriate HAP. 

\subsection{Deployment Scenarios}

To offer qualitative insight about the scalability's gain  of data center-enabled HAP, we compare the system electricity cost; which represents the main operational expenditure source in data centers; according to the deployment scenarios shown in Fig~\ref{fig:3} \cite{hogade2021energy}. In the first scenario, a terrestrial data center  is considered to process all the incoming workload. In the second scenario, some servers are placed in the HAP and most of the servers are hosted in the terrestrial data center. In the third scenario, 4 HAPs collaborate with the terrestrial data center to process the incoming workload. Throughout these simulations, the system electricity cost is evaluated  for the maximum workload arrival rate determined by the HAP flying condition according to the variation of latitudes and days as shown in Fig.~\ref{fig:4a}. {It is worthy to note that the electricity cost of the terrestrial data center includes the cost of the computational energy and the cooling energy. However, the electricity cost of the data center-enabled HAP system captures the saved energy thanks to using the HAP  while considering the wireless transmission energy cost for the communication link from terrestrial base-station to the HAP in addition to cost of utilized terrestrial servers.}

First, by carrying out our own simulations, we notice from  Fig.~\ref{fig:4a} the deployment of a flying data center in the HAP saves substantially the electricity cost of a fully-terrestrial data center with $=12\%$ achieved  when one HAP is deployed and with $=36\%$ when 4 HAPs are deployed respectively for the maximum workload arrival rate.  {Therefore, the electricity costs of terrestrial data centers can be further cut down with  lower arrival rates thanks to HAPs deployment, which reduces the tasks arrival rates to terrestrial data centers.}  Moreover, we notice from  Fig.~\ref{fig:4a} that the highest electricity costs are recorded at the beginning and the end of the year because the lowest solar energy can be harvested during this period in the Northern hemisphere.


\section{Management Agility of Data Center-Enabled HAP}
In this section, we  provide different management techniques    that entail workload management, network management and airship management to enable the desired performance of the data center-enabled HAP.

\subsection{Workload Management}
Incoming workload to data centers are time varying and heterogeneous because they are originated from different applications and serve various users. Therefore, different quality of service  levels are required from the data center enabled-HAP. However, the characteristics of the servers hosted in the HAP are different from those hosted in the terrestrial data center. {For instance, the servers hosted in the HAP are lighter in terms of weight, have less advanced computing features and are limited in number to keep the HAP balanced.} Hence, an agile workload management is imperative to improve the aggregated performance  of the data center-enabled HAP in terms of reliability and delay. 

\begin{figure} [t]
    \centering
   \includegraphics[width=3.4in]{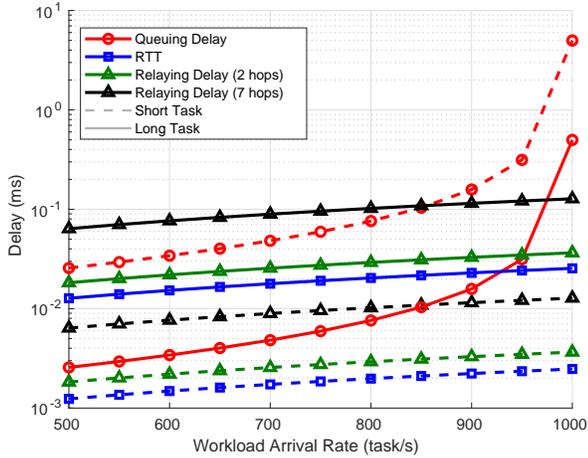}
      \caption{Workload Impact on Delays of Data center-Enabled HAP }
      \label{fig:5}
     \end{figure}

To investigate the data center-enabled HAP delay performance, we consider the task length characteristic of the workload and study its impact on the delay performance versus the workload arrival rate; as shown in Fig.~\ref{fig:5}. For a short task length workload, the transmission delay needed to send the workload to the servers hosted in the HAP and symbolized with round trip time (RTT) is significantly lower compared to the queuing delay experienced in the terrestrial data center. Therefore, the transmission delay can be tolerated  because the workload spends substantial time in the terrestrial servers' queues before processing. However, this observation is not valid when the incoming workload is characterised with a long task length. Specifically, we notice that the transmission delay exceeds the queuing delay for some arrival rates as depicted in Fig.~\ref{fig:5}. 

Another type of delays that needs careful consideration in the multi-HAPs scenario is the relaying delay. We note that the relaying delay represents the time required to relay the workload inter-HAPs. We notice that   the relaying delay is lower than the queuing delay for a 3-HAPs constellation and long task length (2 hops) or 8-HAPs constellation (7 hops) for a short task length workload; as shown in Fig.~\ref{fig:5}. Thus, we conclude that the workload can be relayed between 3 HAPs for long task length and 8 HAPs for short task length without impacting the total delay performance. However, the relaying delay between 3 HAPs exceeds the queuing delay if the workload is characterised with a long task length.

 { The agile workload management is also fundamental for delay-tolerant applications used in far-away areas such as disruption-tolerant networking. The data center-enabled-HAP can be useful in this case when a HAPs' constellation is deployed over different regions to widen the coverage. Accordingly, the delay-tolerant workload   is offloaded to the nearest HAP in the constellation such that the experienced delay is minimized. }

Moreover, an agile workload management that tracks the renewable energy sources is gaining momentum to address the energy concerns of data centers. However, the trade-off between energy demand and renewable energy supply is challenging for different reasons. First, the incoming workload to data centers are time varying and heterogeneous. Second, renewable energy sources are intermittent and depend tightly on weather conditions and geographical locations. 

\subsection{Network Management}
 {The distribution of the computing resources between the HAP and the terrestrial data center engenders supplementary overhead to synchronize and replicate the offloaded data to the HAP.  Therefore, effective network management strategies based on self-organizing approaches should be adopted in the data center-enabled HAP. On the one hand, the control strategy of the data center-enabled HAP's network is fundamental to pool and orchestrate the distributed resources with agility and flexibility and provide them as disaggregated resources in the entire data center-enabled HAP and within a data center-enabled multi-HAPs. Different technologies can be used to implement disaggregation  such as  composable data center infrastructure. This technology is based on the software-defined paradigm to enable the abstraction of the disaggregated resources and their composition/decomposition and management over the HAP or the multi-HAPs and the terrestrial data center.} 

 {On the other hand,  higher bandwidth requirements are imposed on the communication link between the HAP and the terrestrial data center and between the HAPs to embrace the continuously-increasing data  traffic and its control information. In this regard, the terahertz and optical wireless communications of 6G networks will play a key role to handle incoming workload to terrestrial data center when transferred to the stratosphere. Besides, different transmission technologies can be used to enhance the link capacity and optimize its  performance. For instance, massive multiple-input multiple-output (MIMO) offers a high link robustness based on the aggressive  spatial multiplexing  and can be easily implemented on the HAP thanks to its large surface. Massive MIMO can be combined with various access techniques such as orthogonal multiple access  to achieve a high spectral efficiency between the HAP and the terrestrial data center by using maximum ratio combing  and zero-forcing precoding methods.} 

\subsection{HAP Management}
 {Even though HAPs are designed for long duration missions, regular maintenance is required to guarantee the reliable performance of the data center enabled-HAP. First, it is important to supervise if the airship is operating under the prescribed service conditions such as the supported stratospheric turbulence and load. It is also crucial to check the energy storage system of the airship and the electric engines and to monitor the loss percentage of the gas responsible for lifting the airship. }

 {Moreover, it is important to supervise the operation of the data center hosted in the HAP by monitoring the security threats and the non-malicious failures such as  memory leakage and CPU overload. Therefore, secure and dependable mechanisms adapted to the capabilities of the HAP-flying data center should be implemented and periodically updated to offer secure, recoverable and reliable computing services. For instance, the blockchain technology can be deployed in the data center enabled-HAP to secure the data hosted in the HAP. }

 {Besides, advanced redundancy mechanisms can be deployed within  different HAPs in a multi-HAPs constellation setting to fight against the non-malicious failures.  Interestingly, HAPs have the asset to offer an agile maintenance compared to satellites because the possibility of return to Earth is easier.  {However, it is valuable to consider the trade-off between the maintenance costs; which contribute significantly to the operational expenditures; besides the takeoff/landing costs on the one hand and the data center enabled-HAP performance on the other hand. }}

\section{Conclusion and Future Directions}
 {In this article, we shed light on the potential of the data center-enabled HAP to unleash the green computing frontier to the stratosphere. The main goal of this first stage study is to investigate the energy-efficiency and the performance of data center-enabled-HAP. However, it is important to explore  the potential research challenges towards a comprehensive study. From a financial perspective, it is crucial to analyze the capital and operational expenditures of the data center-enabled HAP because they may protract its adoption given the supplementary generated costs of the HAP, the flying servers, the  required regular maintenance. From a technical perspective, it is important to consider the impact of the extreme weather in the stratosphere over the operation of the flying servers and the supported mission duration. It is also essential  to manage the workload in the data center-enabled HAP based on the application requirements and the originating regions. For instance, the multi-HAP scenario can be efficient to handle the workload generated by disruption-tolerant networking in disconnected areas if the necessary network management strategies are adopted. Eventually, it is fundamental to implement artificial intelligence  in workload prediction and network management to improve the performance of the data center-enabled HAP.   However, energy-aware AI algorithms should be applied to conserve the energy-efficiency of the data center-enabled HAP.}


\bibliographystyle{IEEEtran}

\bibliography{bib}

\begin{IEEEbiographynophoto}
{Wiem Abderrahim} (S'14 - M'18) accomplished her undergraduate studies in electrical engineering at the Higher School of Communications of Tunis, Carthage University, Tunisia in 2013. She received her Doctoral Degree in Information and Communication Technologies from the same university in 2017. She worked as a lecturer and then as an adjunct professor 
at the Higher School of Communications of Tunis between 2014 and 2018. Currently, she is a postdoctoral fellow within King Abdullah University of Science and Technology (KAUST), Thuwal, Saudi Arabia. 
\end{IEEEbiographynophoto}
\vspace{-1cm}

\begin{IEEEbiographynophoto}
{Osama Amin} (S'07, M'11, SM'15) received his B.Sc. degree in electrical and electronic engineering from Aswan University, Egypt, in 2000, his M.Sc. degree in electrical and electronic engineering from Assiut University, Egypt, in 2004, and his Ph.D. degree in electrical and computer engineering, University of Waterloo, Canada, in 2010. In June 2012, he joined Assiut University as an assistant professor in the Electrical and Electronics Engineering Department. Currently, he is a research scientist in the CEMSE Division at KAUST, Saudi Arabia. 
\end{IEEEbiographynophoto}
\vspace{-1cm}

\begin{IEEEbiographynophoto}
{Basem Shihada} (SM'12) received the Ph.D. degree in computer science from the University of Waterloo, Waterloo, ON, Canada. He is an Associate and a Founding Professor with CEMSE Division at KAUST, Saudi Arabia. In 2009, he was appointed as a Visiting Faculty Member with the Department of Computer Science, Stanford University, Stanford, CA, USA. 
\end{IEEEbiographynophoto}

\end{document}